\documentclass[10pt,letterpaper,twocolumn]{article} 

\usepackage{ol2SmallCaptions}
\usepackage{comment}
\usepackage[draft]{hyperref}
\usepackage{amsmath}
\usepackage{subfigure}    
\usepackage{empheq}  
\usepackage{graphicx}	
\usepackage{epstopdf}

\begin{document}

\twocolumn[ 

\title{Quantum-correlated photon pairs generated in a commercial 45\,nm
complementary metal-oxide semiconductor microelectronics chip} 


\author{Cale M. Gentry$^{1*}$, Jeffrey M. Shainline$^2$, Mark T. Wade$^1$, Martin J. Stevens$^2$, Shellee D. Dyer$^2$, Xiaoge Zeng$^{1,3}$, Fabio Pavanello$^1$, Thomas Gerrits$^2$, Sae Woo Nam$^2$, Richard P. Mirin$^2$, and Milo\v{s} A. Popovi\'{c}$^1$}
\address{
$^1$Department of Electrical, Computer, and Energy Engineering, University of Colorado Boulder, Boulder, Colorado, 80309, USA\\
$^2$National Institute of Standards and Technology, 325 Broadway, Boulder, Colorado, 80305, USA\\
$^3$Department of Physics, University of Colorado Boulder, Boulder, Colorado, 80309, USA\\
$^*$Corresponding author: cale.gentry@colorado.edu
}

\begin{abstract}
Correlated photon pairs are a fundamental building block of quantum photonic systems.  While pair sources have previously been integrated on silicon chips built using customized  photonics manufacturing processes, these often take advantage of only a small fraction of the established techniques for microelectronics fabrication and have yet to be integrated in a process which also supports electronics.  Here we report the first demonstration of quantum-correlated photon pair generation in a device fabricated in an unmodified advanced (sub-100\,nm) complementary metal-oxide-semiconductor (CMOS) process, alongside millions of working transistors.  The microring resonator photon pair source is formed in the transistor layer structure, with the resonator core formed by the silicon layer typically used for the transistor body. With ultra-low continuous-wave on-chip pump powers ranging from 5 $\mu$W to 400\,$\mu$W,  we demonstrate pair generation rates between 165\,Hz and 332\,kHz using $>$80$\%$ efficient WSi superconducting nanowire single photon detectors.  Coincidences-to-accidentals ratios consistently exceeding 40 were measured with a maximum of 55. In the process of characterizing this source we also accurately predict pair generation rates from the results of classical stimulated four-wave mixing measurements. This proof-of-principle device demonstrates the potential of commercial CMOS microelectronics as an advanced quantum photonics platform with capability of large volume, pristine process control, and where state-of-the-art high-speed digital circuits could interact with quantum photonic circuits.

\vspace{3pt}
\noindent This manuscript describes work of the U.S. government and is not subject to copyright. 
\end{abstract}

 ] 

\section{Introduction}
Quantum photonic systems often consist of relatively large bulk-optical components and can significantly benefit from chip-scale integration \cite{09NPObrien,09Politi,12Tanzilli, 13Metcalf}, similar to how large-scale integration of transistors has revolutionized modern digital electronics. The microelectronics industry, dominated by the complementary metal-oxide-semiconductor (CMOS) technology, remains one of the most successful examples of large-scale integrated systems, benefiting from high-yield and cost-effective fabrication methods while supporting billions of components. Modeling after the success of CMOS, there has been great interest in implementing scalable quantum photonic devices in ``CMOS-compatible" platforms to benefit from proven and reliable fabrication techniques. These silicon photonics processes support high-performance classical devices such as filters \cite{07Popovic,07Xia,13Ong}, switches \cite{08Vlasov,10Lee}, and delay lines \cite{10Melloni, 11Khan, 12Lee} which are essential components of a reconfigurable quantum photonic system. At the heart of such a system must lie a source of photons with non-classical correlations such as entanglement \cite{11Shadbolt}. Quantum-correlated photon pair sources utilizing the relatively large Kerr nonlinearity ($\sim100$ times that of fused silica) of silicon have been proposed \cite{06Lin,11Chen,12Camacho} and demonstrated in silicon wire waveguides \cite{06Sharping,07Takesue,08Harada,12Matsuda}, micro-resonators \cite{09Clemmen,12Azzini,13Engin}, and coupled resonator optical waveguides \cite{12Davanco,13Matsuda,14Takesue}. 
These devices utilize spontaneous four-wave mixing (SFWM) where two pump photons are parametrically converted into a signal and idler photon pair. Recently, time-energy \cite{07Takesue,08Harada,14Takesue,15Grassani,15Suo, 15Wakabayashi} and polarization entanglement \cite{12Matsuda, 15Suo} has been shown between photon pairs generated in silicon sources. Systems have continued to scale to include on-chip interference between multiple integrated photon sources \cite{14Silverstone}, demultiplexing of signal and idler photons \cite{13Kumar,13Collins}, and high-extinction pump rejection \cite{14Harris}. Many of these ``CMOS-compatible" implementations have relied on electron-beam lithography fabrication techniques and often include custom tailored silicon thicknesses that are typical in silicon photonics but are incompatible with advanced CMOS microelectronics, preventing monolithic integration of electronics and quantum photonics on a single chip.   While utilization of CMOS materials and fabrication processes offers certain processing benefits, until now, monolithic integration of quantum photonic sources within a microelectronics platform has not been investigated. 

\begin{figure}[t!]
\includegraphics{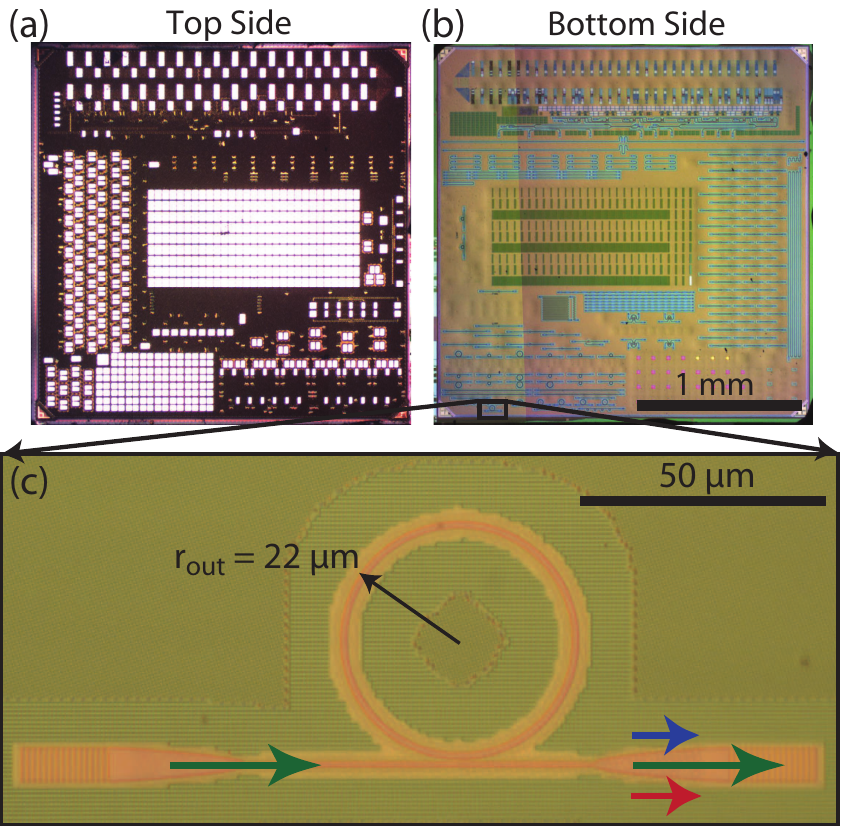}
\caption{\label{fig:Fig1} Optical micrograph of the top (a) and bottom (b) of the CMOS chip with zoom-in (c) of the ring-resonator pair source and grating couplers. }
\end{figure}
Recently, monolithic integration of classical photonics in commercial CMOS processes has been pursued in the context of enabling energy efficient optical interconnects between processors and memory \cite{08Batten} resulting in the demonstration of a chip-to-chip optical link \cite{15Sun}.  The IBM 12SOI 45\,nm CMOS process \cite{07Lee}, utilized for the device in this paper, has proven to be a particularly well-suited platform for integration of photonic devices alongside millions of transistors \cite{12Orcutt} and has enabled control of photonic components by on-chip digital electronics for an optical transmitter and receiver \cite{14Georgas}. High-performance classical photonic components such as 5\,fJ/bit modulators \cite{13Shainline}, record tuning-efficiency filters \cite{14Wade}, and highly efficient fiber-to-chip grating couplers \cite{15Wade} have all been demonstrated in commercial CMOS, but integration of parametric sources has not been explored until now. Furthermore, microelectronic circuits in the 45\,nm SOI CMOS process used here have been shown to operate at cryogenic temperatures compatible with quantum systems \cite{10Vernik}, opening the possibility to monolithically integrated single-chip quantum optical processors. In this paper, we demonstrate the first source of quantum-correlated photon pairs directly integrated in an unmodified advanced CMOS process.  

\begin{figure}[t!]
\includegraphics{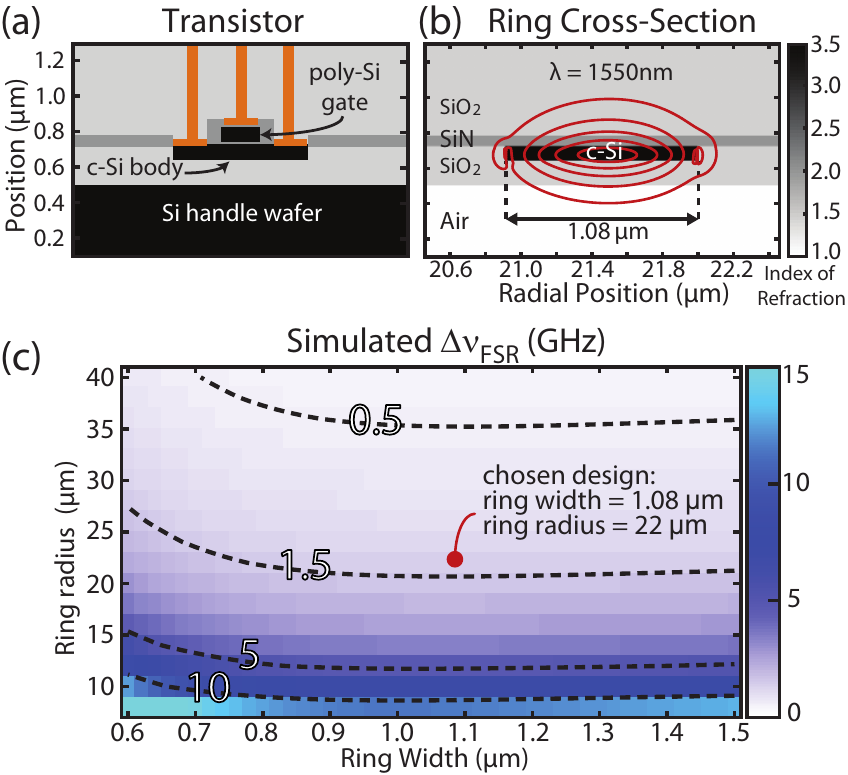}
\caption{\label{fig:Fig2} (a) Schematic of a typical transistor composed of a crystalline silicon (c-Si) body and a polysilicon gate. (b) Cross-section illustration of the microring-resonator pair source showing how the sub-100\,nm c-Si transistor body layer can be used to confine light after removal of the Si handle wafer. The fundamental resonator mode contours are superimposed in red to illustrating how the majority of the modal field extends into the cladding. (c) Simulated difference in FSR $\Delta\nu_\mathrm{FSR}$ at 1550\,nm due to dispersion with chosen design predicted to be a negligible 1.4\,GHz.}
\end{figure}

\begin{figure}[t!]
\includegraphics{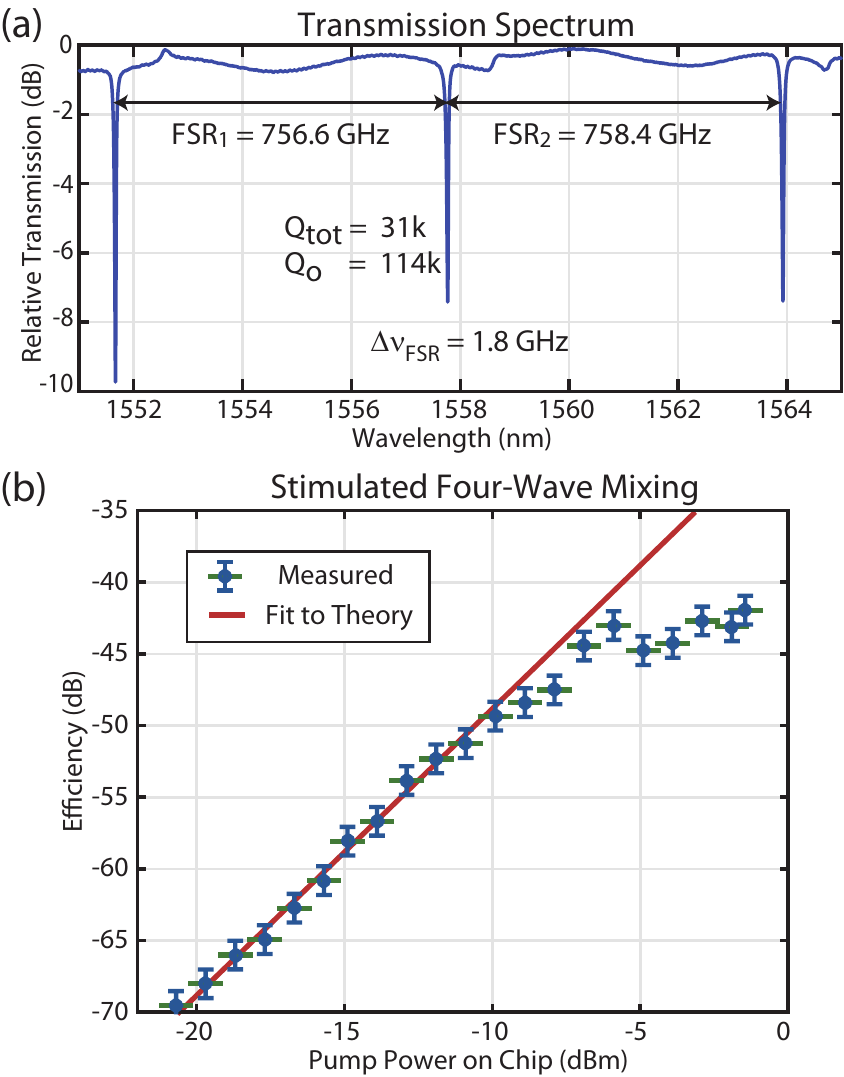}
\caption{\label{fig:Fig3} (a) Passive wavelength sweep of the three interacting resonances showing a difference in FSR of 1.8\,GHz and intrinsic quality factor of 114,000. (b) Measured stimulated four-wave mixing efficiency with fit to Eq. (\ref{stFWMEff}). Horizontal (vertical) error bars correspond to uncertainty in input (output) coupling. Deviation from theory at higher pump powers are result of parasitic nonlinear and thermal effects.}
\end{figure}

\section{Device Design}
As a photon pair source we use a microring-resonator fabricated in the crystalline silicon (c-Si) CMOS layer typically used for the body of a transistor [Fig. \ref{fig:Fig2}(a)]. The silicon handle wafer is removed to provide confinement of the optical mode \cite{12Orcutt}.  The sub-100\,nm thick \cite{NDA} c-Si guiding layer results in a significant portion of the optical mode extending into the SiN and SiO$_2$ cladding as shown in Fig. \ref{fig:Fig2}(b). This not only confines a small fraction of the modal power within the silicon core, the dominant Kerr medium, but also prevents the design of a resonator with zero group velocity dispersion. Four-wave mixing is greatly enhanced by the presence of a large density of photonic states which conserve energy and momentum. When a degenerate pump beam is tuned on resonance, adjacent resonances are intrinsically momentum matched but dispersion results in a difference between adjacent free-spectral ranges (FSR) introducing an energy mismatch \cite{14FiOGentry}.  This effect of dispersion becomes negligible if the difference in FSRs,  $\Delta\nu_\mathrm{FSR}$, is significantly smaller than the linewidths of the resonances involved. Simulations of $\Delta\nu_\mathrm{FSR}$ are shown in Fig. \ref{fig:Fig2}(c) for a pump wavelength near 1550\,nm for various ring widths and radii.  A ring width of 1.08\,$\mu$m and outer radius of 22\,$\mu$m with a predicted difference in FSR of 1.4\,GHz is chosen in order to obtain a small mode volume while also supporting a suitably small $\Delta\nu_\mathrm{FSR}$. A transmission spectrum [Fig. \ref{fig:Fig3}(a)] of the CMOS foundry-fabricated device shows a difference in FSRs of 1.8\,GHz centered around a pump resonance near 1558\,nm.  The 0.4\,GHz difference between measured and simulated values is likely due to uncertainty in refractive indices and fabricated dimensions. Fitting the passive resonance gives a total quality factor, $Q_\mathrm{tot}$, of $31,000$, with an intrinsic quality factor of $Q_\mathrm{o}$ = $114,000$ due to intrinsic losses such as linear absorption and roughness loss and an external quality factor of $Q_\mathrm{ext}$ = $43,000$ due to coupling to the waveguide bus. The total $Q$ corresponds to a linewidth of 6.2\,GHz, which is significantly greater than the measured 1.8\,GHz difference in FSRs.  The resonator is over-coupled to provide higher tolerance to dispersion while also providing a higher escape efficiency for generated photon pairs. 

\section{Stimulated Four-wave Mixing}
 In order to characterize the photon pair source, we first measure stimulated four-wave mixing (FWM) where a seed laser is used in addition to the pump to stimulate the four-wave mixing process. The efficiency of classical FWM is commonly defined as the ratio of idler power in the output bus to seed power in the input bus. Measuring FWM with pump powers ranging from -21\,dBm to -2\,dBm results in efficiencies of -70\,dB to -42\,dB, respectively, as shown in Fig. \ref{fig:Fig3}(b).  The efficiencies with pump power below -10\,dBm follow an expected quadratic dependence on pump power. At higher pump powers, parasistic nonlinearities such as two-photon absorption (TPA), free-carrier absorption (FCA), and self- and cross-phase modulation, in addition to thermal heating of the resonator due to linear absorption, result in a deviation from the quadratic trend.
Assuming that the pump and seed lasers are both placed on resonance, in the limit of negligible nonlinear loss, the efficiency of stimulated FWM is well described by \cite{14OLGentryoO, 15Zeng}
\begin{eqnarray}
\eta_\mathrm{stim} = 
P_\mathrm{p}^2 \omega^2 \beta_{\mathrm{fwm}}^2 \left(\frac{2r_{\mathrm{ext}}}{r_{\mathrm{tot}}^2}\right)^3  \frac{2r_{\mathrm{ext}}}{(2\pi \Delta \nu_\mathrm{FSR})^2 + r_{\mathrm{tot}}^2} \label{stFWMEff}.
\end{eqnarray}
Here, $\Delta \nu_\mathrm{FSR}$ is the difference in adjacent FSRs due to dispersion, $P_\mathrm{p}$ is the pump power in the waveguide at optical angular frequency $\omega$, and $\beta_\mathrm{fwm}$ is a parameter containing the nonlinearity of silicon and is inversely proportional to the mode volume. The decay rates $r_{\mathrm{tot}}$ (total energy amplitude decay rate due to all mechanisms) and $r_{\mathrm{ext}}$ (decay rate due to coupling to waveguide only) are related to the measured quality factors $Q_\mathrm{tot} = {\omega}/{2r_\mathrm{tot}}$  and $Q_\mathrm{ext} = {\omega}/{2r_\mathrm{ext}}$.  The decay rates in the resonator presented here are treated as equal for the pump, signal, and idler resonances.
Fitting the measured efficiencies in the region of pump powers before parasitic nonlinearities are present (below -10\,dBm) to Eq. (\ref{stFWMEff}) we find $\beta_\mathrm{fwm} = 5.32\times10^6$ J$^{-1}$.  The $\beta_\mathrm{fwm}$ parameter is related to the Kerr nonlinearity, denoted $n_2$, of the silicon core by 
\begin{eqnarray}
\beta_\mathrm{fwm} = \frac{n_2 \, c}{n_\mathrm{Si}^2 V_\mathrm{eff}},
\end{eqnarray}
where $c$ is the speed of light, $n_\mathrm{Si}$ is the refractive index of silicon, and $V_\mathrm{eff} = 18.5\, \mu \mathrm{m}^3$ is an effective mode volume \cite{14OEZeng} calculated using a numerical modesolver \cite{03Popovic}. 
Here, the fitted $\beta_\mathrm{fwm}$ corresponds to $n_2 = 3.96\times 10^{-14}$\,cm$^2$/W, well within the uncertainties of previously measured values for crystalline silicon \cite{03Dinu,07Bristow}.

\begin{figure}[t!]
\includegraphics{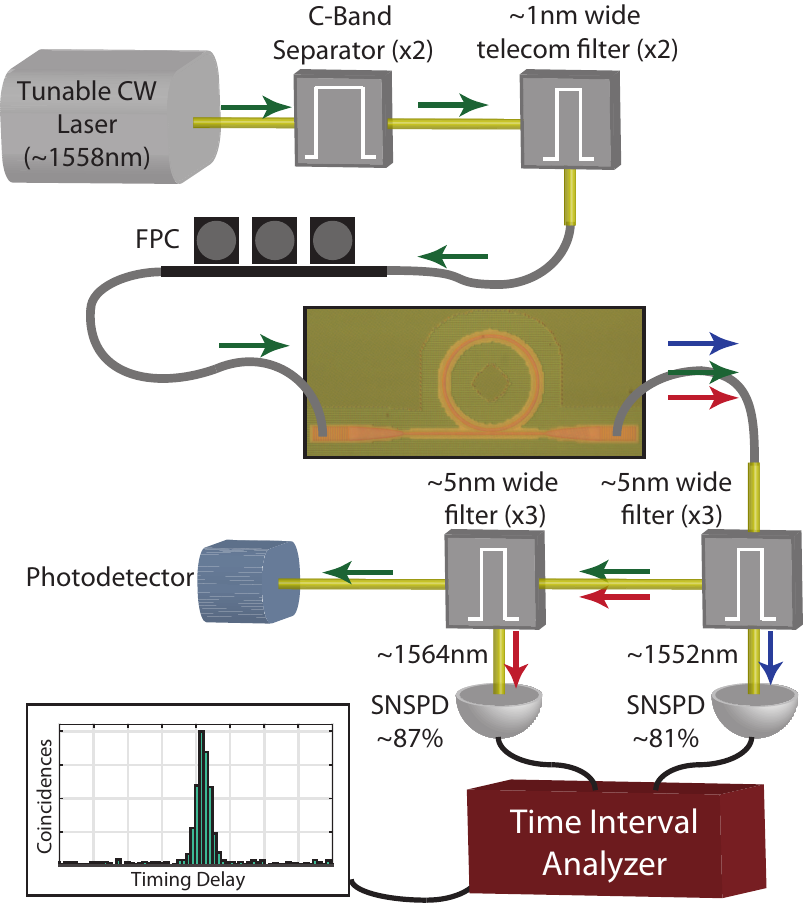}
\caption{\label{fig:Fig4} Simplified schematic of the pair generation measurement. The pump light is passed through two C-band (1530-1565\,nm) separators and two telecom filters to eliminate noise from the laser. A fiber polarization controller (FPC) is used to optimize coupling efficiency.  The signal and idler photons are then filtered individually and sent to high efficiency superconducting nanowire single photon detectors (SNSPDs).  A time interval analyzer is then used to count coincidences. The pump power is monitored by a classical photodetector to ensure the pump light is on resonance. }
\end{figure}

\section{Photon Pair Generation}
Spontaneous four-wave mixing is tested by performing coincidence measurements between generated photon pairs.  A schematic of the experimental setup is illustrated in Fig. \ref{fig:Fig4}. A continuous-wave (CW) telecom pump laser is passed through a series of two C-band (1530-1565\,nm) separators and two $\sim$1\,nm wide telecom bandpass filters to remove laser spontaneous emission noise at the signal and idler wavelengths. The pump is subsequently coupled to the input waveguide via a grating coupler and tuned to the pump resonance near 1558\,nm . The generated photon pairs are coupled from the output grating and then individually filtered by cascaded telecom filters with an estimated 180\,dB total isolation from the pump. The signal and idler photons are sent to $81\%$ and $87\%$ efficient WSi superconducting nanowire single photon detectors (SNSPDs) \cite{13Marsili}, respectively. While microring-resonator sources of photon pairs generally generate a comb of signal-idler pairs since FSRs multiple mode orders away from the pump resonance are often also phase-matched, we use $\sim$5\,nm bandwidth telecom filters at the signal ($\sim$1552\,nm) and idler ($\sim$1564\,nm) wavelengths to ensure measurement of pairs only at the FSRs immediately adjacent to the pump.  A time interval analyzer records counts as a relative time delay between the two detectors.  

\begin{figure}[b!]
\includegraphics{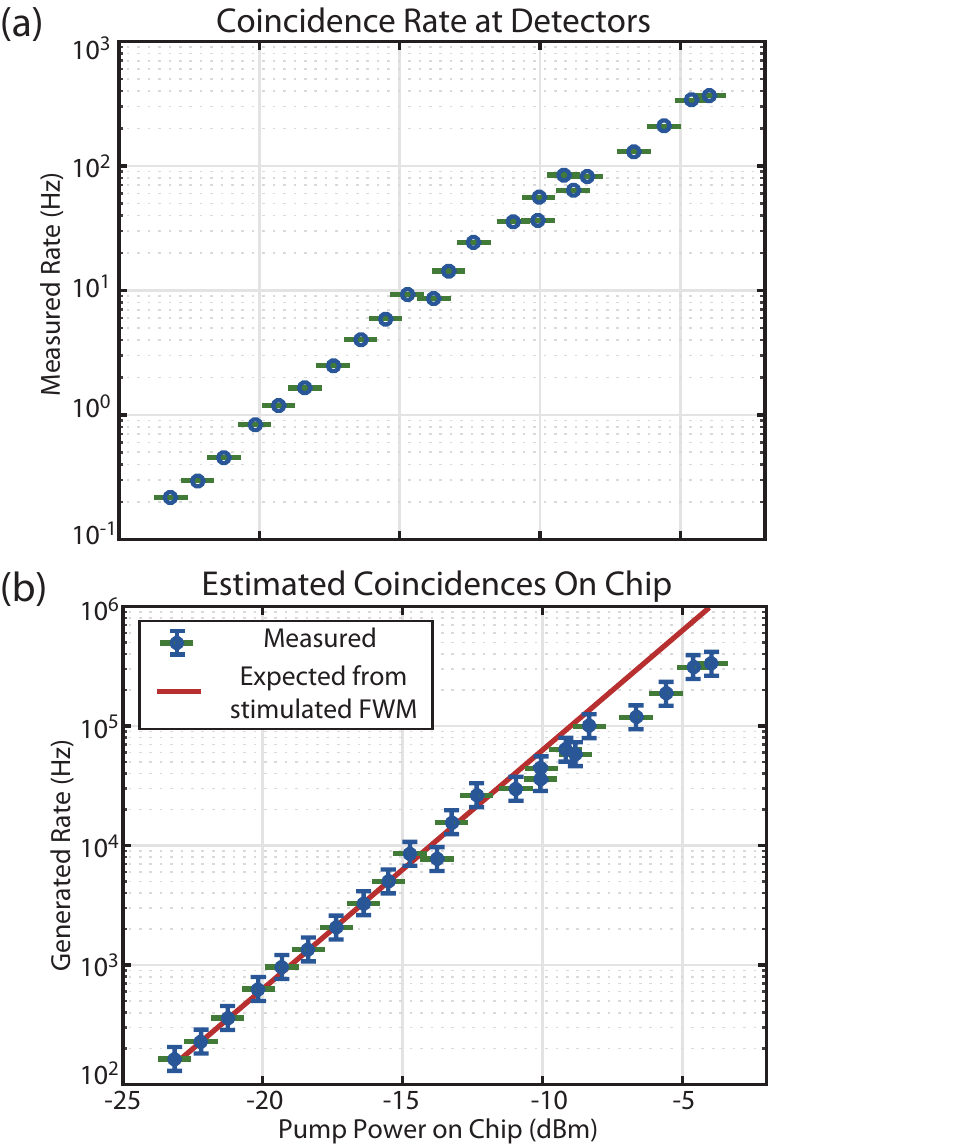}
\caption{\label{fig:Fig5}(a) Coincidence rate of photon pairs at the detectors with the estimated rate on-chip (b) after subtracting out losses to the detectors.  The solid line is the rate expected based on the stimulated four-wave mixing measurements in Fig. \ref{fig:Fig3}(b). }
\end{figure}
From here, a coincidence rate versus estimated pump power in the waveguide is measured as shown in Fig. \ref{fig:Fig5}(a).  Measurements at each pump power are performed with a sufficiently long integration time to accumulate the same number (approximately 200 counts) of coincidences at the zero-delay time bin. The integration times ranged from 2\,s at the highest pump power to 1\,h at the lowest pump power. Fig. \ref{fig:Fig5} (b) shows the estimated pair generation rate on chip after detection efficiency and losses from the output waveguide to the detectors ($\sim15$\,dB)  are taken into account. Measurements of grating coupler loss are performed immediately after each coincidence measurement in order to account for potential drift in the coupling to the waveguide. The uncertainty of individual grating coupler loss is visible in the error bars in Fig. \ref{fig:Fig5} because only the combination of both input and output gratings, which we assume to be identical, can be directly measured. 

\begin{figure}[t!]
\includegraphics{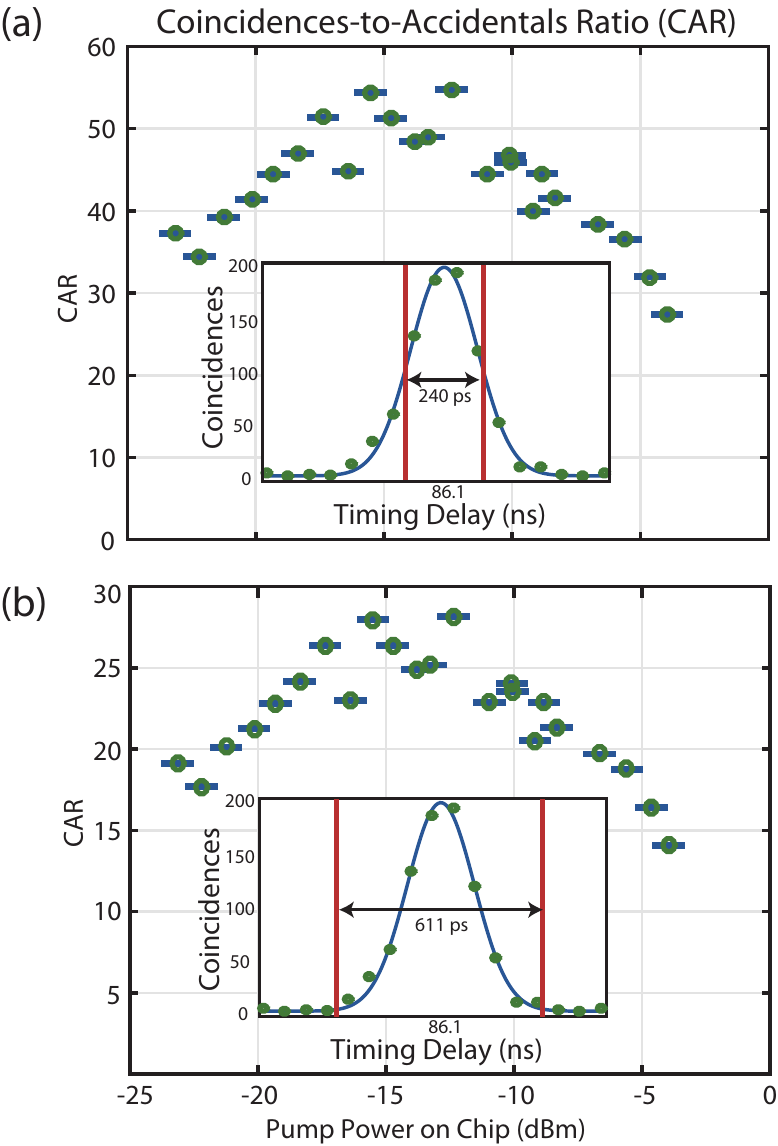}
\caption{\label{fig:Fig6} Coincidences-to-accidentals ratio (CAR) for various pump powers where the coincidence peaks were fit to a Gaussian function (insets) and delay windows selected at the (a) full-width-half-maximum and (b) $\pm$3 standard deviations.}
\end{figure}
Pair generation rates were measured over three orders of magnitude with on-chip rates ranging from 165\,Hz to 332\,kHz. The rate of photon pairs in the output waveguide can be described by \cite{15Vernon}
\begin{eqnarray}
I_\mathrm{coinc} =  \eta_\mathrm{esc} P_\mathrm{p}^2 \omega^2 \beta_\mathrm{fwm}^2 \left(\frac{2 r_\mathrm{ext}}{r_\mathrm{tot}^2}\right)^2 \times \nonumber \\ 
\frac{4 r_\mathrm{ext}}{(2 \pi \Delta\nu_\mathrm{FSR})^2+(2r_\mathrm{tot})^2}, \label{spFWM}
\end{eqnarray}
where $\eta_\mathrm{esc}$ is an escape efficiency for photons generated in the ring defined as $r_\mathrm{ext}/r_\mathrm{tot}$.  The escape efficiency takes into account the photon pairs generated within the cavity where one of the photons is lost due to loss mechanisms such as sidewall roughness scattering and absorption. Use of the fitted $\beta_\mathrm{fwm}$ from the classical FWM measurements to predict pair generation rate provides excellent agreement with the measured data, as seen in Fig. \ref{fig:Fig5}(b). Similar to the case of classical measurements, at higher pump powers the pair rate experiences a deviation from theory due to parasitic nonlinear \cite{13Engin,13Husko,13Helt} and thermal effects.  

Coincidence measurements also provide a coincidences-to-accidentals ratio (CAR), often used as a figure of merit characterizing the noise of a photon pair source. The coincidence peak has a finite width resulting from a combination of timing jitter of the detectors ($\sim$105\,ps and $\sim$130\,ps for the signal and idler, respectively) and the temporal width of the photon pairs determined by the linewidth of the resonator. Here, we perform a Gaussian fit to the coincidence peak and use the full-width-half-maximum (FWHM) as the delay window [inset of Fig. \ref{fig:Fig6}(a)] for comparison to previously demonstrated photon pair sources.  The measured CARs are shown versus pump power in Fig. \ref{fig:Fig6}(a).  CARs were consistently measured greater than 40 with a maximum CAR of 55 at a pump power of -12.4\,dBm.  The choice of FWHM as the timing window is somewhat arbitrary \cite{13Engin,09Dyer}, though useful for comparison to other demonstrations in the literature. A larger coincidence window results in a larger pair rate but lower CAR, while a smaller coincidence window results in a higher CAR at the sacrifice of pair rate. Since the measured coincidence rates in Fig. \ref{fig:Fig5} include all true coincidences, we provide the corresponding CARs in Fig. \ref{fig:Fig6}(b) where the timing window is chosen to cover $\pm$3 standard deviations ($\sim611$\,ps) of the fitted Gaussian. We note, the larger window results in about 50$\%$ lower CARs compared to using the FWHM.  Since SNSPDs typically display negligible intrinsic dark counts ($<$1\,count per second \cite{13Marsili}), the CAR is primarily limited by background counts likely originating from spontaneous Raman emission generated in the cladding, leakage of the pump light through the filters, and room-temperature thermal radiation reaching the detectors. 

\section{Discussion}

Despite a device geometry limited by implementation in a 45\,nm-node CMOS microelectronics process, the pair source presented here demonstrated high generation rates up to 332\,kHz and CARs exceeding 50.  These pair rates and CARs are on the same order as many custom fabricated Si sources \cite{09Clemmen, 12Azzini, 12Davanco, 13Kumar,14Harris}.  CARs can be improved by increased rejection of background emission from the pump laser and by the use of a pulsed pump \cite{09Dyer} or an on-chip modulator \cite{13Shainline} to control pair generation time windows. While the CMOS process does not support a completely dispersionless single-ring design, we showed that the dispersion could be minimized, resulting in a mere 1.8\,GHz difference in adjacent FSRs.  Equations (\ref{stFWMEff}) and (\ref{spFWM}) can be used to quantify the effect of dispersion as a decrease in efficiency of $\sim$\,14.6$\%$ and  $\sim$\,7.5$\%$ for stimulated and spontaneous four-wave mixing, respectively. 

The relationship between stimulated and spontaneous four-wave mixing has been a subject of recent investigation \cite{12Helt,12AzziniOL} as it is useful to determine the effectiveness of predicting pair generation rates from classical FWM measurements. In fact, classical measurements have recently been demonstrated for fast and efficient characterization of entangled-photon sources \cite{15Rozema}. In the limit of no dispersion, Equations (\ref{stFWMEff}) and (\ref{spFWM}) give a simple formula relating pair generation rate to seeded four-wave mixing efficiency of
\begin{eqnarray}
I_\mathrm{coinc} =  \frac{\eta_\mathrm{esc}}{2} \left(\frac{2 r_\mathrm{ext}}{r_\mathrm{tot}^2}\right)^{-1} \eta_\mathrm{stim}.
\end{eqnarray}
Until recently \cite{15Vernon}, a complete description of SFWM in lossy microcavities had not been available, leading to a significant discrepancy in measured pair rates and theoretical predictions \cite{09Clemmen}.  Here we accurately predict the SFWM-based pair rates from classical FWM measurements of a microresonator. These results both confirm the findings of \cite{12AzziniOL} and extend them by predicting non-classical correlations via coincidence rates (rather than generated optical powers) in the more general case of arbitrary coupling strength and measurable dispersion. 

Pair generation rates near 165\,Hz with on-chip pump powers as low as 4.8\,$\mu$W were demonstrated, which is, to the best of our knowledge, the lowest pump power used to produce photon pairs in silicon. This was primarily enabled by the first use of highly efficient SNSPDs for coincidence measurements of pairs generated in a Si microring source, despite the significant loss from the output grating coupler to the detectors.  These losses can be greatly reduced in future implementations by integration with on-chip filters and highly efficient grating couplers which have already been demonstrated in this process \cite{15Wade}, potentially enabling heralded single photon sources with high heralding efficiency, necessary to compete with parametric down-conversion sources \cite{14Dixon}.

\section{Conclusion}
We have demonstrated the first source of quantum-correlated photons in an advanced microelectronics process, opening the door to the integration of quantum states of light with photonic circuits and high-speed digital electronics on a single chip. In addition, this is the first measurement of photon pairs from a silicon microresonator using highly efficient ($>$80$\%$) single photon detectors allowing record low pump power pair generation.  This work also provides an accurate prediction of pair generation rates of a microresonator photon pair source from classical four-wave mixing measurements.  The results presented here are a proof-of-principle demonstration that quantum sources of light can be integrated within a standard CMOS platform. Combined with previously demonstrated classical components, the CMOS process may provide an attractive platform for quantum photonic circuits controlled by state-of-the art electronics, even at cryogenic temperatures \cite{10Vernik}.  For example, previously demonstrated on-chip classical detectors could monitor the power of a pump laser from the through-port of a microresonator photon pair source like that demonstrated in this paper, while digital circuits could actively compensate for drift in the resonance frequency caused by environmental factors such as temperature fluctuations, thereby enabling a highly stable source of photon pairs. One can also imagine a reconfigurable feed-forward system \cite{11Shadbolt,07Prevedel} where detection of a heralding signal photon can be processed by digital logic on chip and subsequently route the corresponding idler photon for a specific operation. The results presented here extend beyond the simple integration of photonic circuits in ``CMOS compatible'' materials or widely used customized photonics processes by implementing a source directly in an unmodified commercial CMOS process. 

\section*{Funding Information}
This work was supported by a CU-NIST Measurement Science and Engineering (MSE) Fellowship and by the Office of Naval Research Grant N000141410259. 

\section*{Acknowledgments}
We would like to thank Gil Triginer Garc\'{e}s, Scott Glancy, L. Krister Shalm and Timothy Bartley for helpful discussions related to this work.


\end{document}